\documentclass[
showpacs,
preprintnumbers,
amsmath,
amssymb,
prd,
nofootinbib,
superscriptaddress,
aps,
onecolumn
]{revtex4-1}
\usepackage{amsmath}
\usepackage{subfigure}
\usepackage{amsfonts}
\usepackage{hyperref}
\usepackage{amssymb}
\usepackage{graphicx}
\graphicspath{ {figs/} }
\usepackage[latin1]{inputenc}
\usepackage[english]{babel}
\usepackage{epsfig}
\usepackage{color,xcolor}

\definecolor{purple}{rgb}{1,0,1}
\definecolor{lime}{HTML}{A6CE39} 

\def\eqn#1{\eq\eqref{#1}}
\def\rf{\eqref}

\def\qq{\qquad}

\def\lal{&&\ {}}
\def\eq{Eq.\,}

\def\beq{\begin{equation}}
\def\eeq{\end{equation}}
\def\bear{\begin{eqnarray}}
\def\bearr{\begin{eqnarray} \lal}
\def\ear{\end{eqnarray}}
\def\earn{\nonumber \end{eqnarray}}
\def\nn{\nonumber\\ {}}

\def\nnn{\nonumber\\ \lal }

\def\yyy{\\[5pt] \lal }

\def\const{{\rm const}}
\def\diag{\,{\rm diag}\,}

\def\then{\ \Rightarrow\ }
\def\d{\partial}

\begin{document}

\title{On black bounce space-times in non-linear electrodynamics}

\author{G. Alencar}
\email{geova@fisica.ufc.br}
\affiliation{Department of Physics, Universidade Federal do Cear\'a (UFC), Campus do Pici, Fortaleza - CE, C.P. 6030, 60455-760 - Brazil}

\author{Kirill A. Bronnikov}
\email{kb20@yandex.ru}
\affiliation{Center of Gravitation and Fundamental Metrology, VNIIMS, Ozyornaya ulitsa 46, Moscow 119361, Russia}
\affiliation{Peoples' Friendship University of Russia, ulitsa Miklukho-Maklaya 6, Moscow, 117198, Russia}
\affiliation{National Research Nuclear University ''MEPhI'', Kashirskoe shosse 31, Moscow 115409, Russia}

\author{Manuel E. Rodrigues}
\email{esialg@gmail.com}
\affiliation{Faculdade de Ci\^{e}ncias Exatas e Tecnologia, 
Universidade Federal do Par\'{a}\\
Campus Universit\'{a}rio de Abaetetuba, 68440-000, Abaetetuba, Par\'{a}, Brazil}
\affiliation{Faculdade de F\'{\i}sica, Programa de P\'{o}s-Gradua\c{c}\~ao em 
F\'isica, Universidade Federal do 
Par\'{a}, 66075-110, Bel\'{e}m, Par\'{a}, Brazil}

\author{Diego S\'{a}ez-Chill\'on G\'{o}mez}
\email{diego.saez@uva.es}
\affiliation{Department of Theoretical, Atomic and Optical Physics, Campus Miguel Delibes,
University of Valladolid UVA, Paseo Bel\'{e}n, 7, 47011 Valladolid, Spain}
\affiliation{Department of Physics, Universidade Federal do Cear\'a (UFC), Campus do Pici, Fortaleza - CE, C.P. 6030, 60455-760 - Brazil}


\author{Marcos V. de S. Silva}
\email{marco2s303@gmail.com}
\affiliation{Department of Physics, Universidade Federal do Cear\'a (UFC), Campus do Pici, Fortaleza - CE, C.P. 6030, 60455-760 - Brazil}
\affiliation{Faculdade de F\'{\i}sica, Universidade Federal do Par\'{a}, Campus Universit\'{a}rio de Salin\'{o}polis,
68721-000, Salin\'opolis, Par\'{a}, Brazil}

\begin{abstract}

One of the main issues in gravitation is the presence of singularities in the most common space-time solutions of General Relativity, as the case of black holes. A way of constructing regular solutions that remove spacelike singularities consists in implement a bounce on such space-time, leading to what is usually known as black bounce space-times. Such space-times are known to describe regular black holes or traversable wormholes. However, one of the main issues lies on reconstructing the appropriate source that leads to such a solution. In this paper, a reconstruction method is implemented to show that such types of metrics can be well accommodated in non-linear electrodynamics with the presence of a scalar field. Some of the most important black bounces solutions are reconstructed in this framework, both in 3+1 as in 2+1 dimensions. For the first time in the literature, these solutions have an electrically charged source of matter from non-linear electrodynamics.
Specific features are indicated that distinguish electric sources from magnetic ones, 
	  previously found for the same space-times. 
\end{abstract}
\maketitle

\section{Introduction}
\label{intro}

Space-time singularities arise commonly in Einstein's theory of General Relativity (GR) under very
natural assumptions \cite{Penrose:1964wq}, which are known to push the validity limits of the theory. Nevertheless, those solutions containing spacelike or timelike singularities, as black holes or Friedmann-Lema\^itre-Robertson-Walker space-times, seem to describe accurately well real physical systems. In particular, black holes are known to be the final state of critical collapse for those massive stars that, after the nuclear fuel is over, do not get stabilized as the degeneration pressure is not sufficient to hold the star \cite{Joshi}. In GR, such collapsing objects form remediless a spacelike singularity and eventually an event horizon that hides the geodesics incompleteness for external observers. Hence, over the last decades a great effort has been followed on understanding such objects as the edges of GR. In addition, these solutions have also attracted a lot of attention since 
the first detection of gravitational waves which are known to be generated by the coalescence of binary systems formed by black holes and/or neutron stars \cite{LIGOScientific:2016aoc}. Also, reconstruction of the images surrounding supermassive black holes at the centers of M87 galaxy 
and the Milky Way have provided new ways of analyzing such objects in the search for GR extensions and new physics \cite{EventHorizonTelescope:2019dse,EventHorizonTelescope:2022wkp}.

Then, one may say that this is the starting point of a new era in gravitational physics, when precise tests at the strong-field regime of gravity can be performed. In this sense, going beyond GR (and/or beyond the Standard Model of Particles) means that one may leave behind the family of solutions that describe every black hole in GR, the so-called Kerr family, in which all the features of a particular black hole are characterized by just three parameters: its mass, angular momentum and charge, as states the no-hair theorem. However, beyond standard physics, other types of solutions are found, which might describe black holes as well but also other more exotic objects as wormholes, gravastars or boson stars, among many others (for a taxonomy on compact objects see \cite{Cardoso:2019rvt}). Nevertheless, the search for black hole-like solutions that remain regular everywhere is being one 
of the main scenarios explored currently. While the first work pointing to such possibility now is half 
a century old \cite{bardeen:1968non}, many others have been published recently, specially in the aim of exploring theories beyond GR or high dimensions \cite{Lan:2023cvz,Bambi:2023try,Olmo:2015axa,Olmo:2012nx,Bueno:2024dgm,Karakasis:2023hni}. In this sense, some models explore the possibility of removing the space-time region containing the singularity and matching it to another regular region through --- generally --- a wormhole, while some others push the geodesics away 
from the singularity to an infinite value of its affine parameter (see, for instance, 
\cite{Guerrero:2020uhn}). 

Another simple way of regularizing some well-known solutions from 
GR was proposed in \cite{Simpson:2018tsi}, where basically the radius of the 2-sphere is forced 
to be bounded from below, such that the inner region of such a static object reaches a minimum size.
In a globally static case, a wormhole throat connects the original region with another exterior 	
	and regular region, making, as a whole, a traversable wormhole. In cases where the solution 	
	contains an event horizon, and the minimum size is reached in a nonstatic region beyond it, 
	such a minimum is called a black bounce (BB), and the whole configuration then has the nature 
	of a regular black hole. Sometimes, all solutions regularized in this way are called ``BB solutions.'' 
	
Explicit wormhole solutions have been widely explored in the literature, despite their speculative character that is sometimes observed just as an academic training exercise 
\cite{Morris:1988cz}. However, in general, wormholes provide the best way of removing the inner singularity from black holes, as shown in some previous works 
\cite{Olmo:2015dba,Simpson:2021vxo,Visser:1995cc}. In addition, the possibility of the existence 
of such type of solutions have been intensively explored, not only from the theoretical point of view 
but also regarding some possible observational effects, as gravitational lensing \cite{Islam:2021ful, Nascimento:2020ime, Tsukamoto:2020bjm,Ghosh:2022mka,AbhishekChowdhuri:2023ekr} or the images formed when the object is surrounded by an accretion disk \cite{Guerrero:2021ues,Guerrero:2022qkh,Olmo:2021piq}.   

In this paper, we focus on the analysis of several BB space-times and particularly explore the way they can be reproduced in the framework of GR with sources formed by non-linear electrodynamics (NED) and a canonical or phantom scalar field. A significant issue arises when trying to obtain a consistent source that generates such BB solutions since non-canonical sources are necessary \cite{lobo20}. In this sense, NED seems to provide a way of reconstructing BB solutions (see Refs.
\cite{Guerrero:2020uhn,Olmo:2011ja,Olmo:2011np,Olmo:2013gqa,Bronnikov:2022ofk,rodrigues23,
rahul22,Bronnikov:2023aya,Lima:2023arg,canate22}). Moreover, such solutions are not restricted to spherically symmetric space-times but also include regular black strings, i.e. space-times with cylindrical symmetry \cite{Bronnikov:2023aya,Lima:2023arg,Lima:2022pvc,Lima:2023jtl}. Hence, here we review some of these solutions, both spherically symmetric and cylindrically symmetric ones, and implement 
a reconstruction procedure to obtain the corresponding NED Lagrangian together with the corresponding scalar field potential and its canonical or phantom nature depending on the sign 
of its kinetic term. Moreover, regular black hole solutions are also explored in $2+1$ dimensions, including the regularized version of the Ba\~nados-Teitelboim-Zanelli (BTZ) black hole that forms 
a case of BB \cite{Furtado:2022tnb}. Finally, the Null Energy Condition (NEC) is also analyzed in a general way applicable to any of the space-times considered in the paper. 

An important feature of the present study is that we here consider NED sources in the form
    of an electric field instead of previously used magnetic ones 
    \cite{rodrigues23,rahul22,Bronnikov:2023aya,Lima:2023arg,canate22} and attract attention 
    to an important property of electric NED solutions, that they quite generically require 
    different NED Lagrangians in different parts of space. This happens if the transformed
    electromagnetic invariant $P$ (see \eqn{Pscalar} below) is not a monotonic function of 
    the standard invariant $F = F_{\mu\nu} F^{\mu\nu}/4$ 
    (see \cite{Bronnikov:2022ofk, kb01-NED, kb18-NED}  for details), 
    and we come across this phenomenon in some BB solutions analyzed in this study. 

The paper is organized as follows: Section \ref{NED} introduces the action containing a NED Lagrangian and a scalar field. Section \ref{Spherical} is devoted to spherically symmetric BB, 
while Section \ref{cyllindrical} deals with black string space-times. Black bounces in $2+1$ 
dimensions are explored in Section \ref{21dimensions}. In Section \ref{EN}, the energy conditions 
are analyzed. Finally, Section \ref{conclusions} gathers the conclusions of the paper.

\section{Non-linear electrodynamics in the presence of a scalar field}
\label{NED}

 Along this paper, we are considering the Hilbert-Einstein action in the presence of a scalar field 
 and a NED Lagrangian, which can be expressed as follows:
\beq
    S=\int \sqrt{-g}d^4x\left[\frac{R}{2\kappa^2 } +\epsilon g^{\mu\nu}\d_\mu \phi\d_\nu \phi
    		-V(\phi) - L(F)\right]\ ,    			\label{Action}
\eeq
  where $g$ is the determinant of the metric $g_{\mu\nu}$, $\kappa^2=8\pi$, $\phi$ is the scalar field, $V(\phi)$ the scalar field potential, $L(F)$ is the NED Lagrangian, 
  $F=\frac{1}{4}F^{\mu\nu}F_{\mu\nu}$, and $F_{\mu\nu} = \d_\mu A_\nu - \d_\nu A_\mu$ is the 
  electromagnetic field tensor. In addition, $\epsilon=\pm1$ depending on whether the scalar field is canonical ($+$) or phantom ($-$).
  
The corresponding field equations are obtained by varying the action \eqref{Action} with respect to $\phi$, $A_\mu$, and $g^{\mu\nu}$, leading to:
\bearr
		\nabla_\mu \left[L_F F^{\mu\nu}\right]
			=\frac{1}{\sqrt{-g}}\d_\mu \left[\sqrt{-g}L_F F^{\mu\nu}\right]=0\ ,
														\label{eq-F}
\\  \lal
      2\epsilon \nabla_\mu \nabla^\mu\phi =
		-  \frac{dV(\phi)}{d\phi}\ ,                      \label{eq-phi}
\\  \lal
     G_{\mu\nu}=R_{\mu\nu} - \frac 12 g_{\mu\nu} R = \kappa^2\left(T_{\mu\nu} ^{\phi} + T_{\mu\nu} ^{EM}\right)\ ,				\label{eq-Ein}
\ear
  where $L_F=d L/d F$ whereas $T_{\mu\nu}^{\phi}$ and $T_{\mu\nu}^{EM}$ are the stress-energy tensors for the 
  scalar field and electromagnetic field, respectively, which are given by:
\bearr                   \label{SET-F}
    T_{\mu\nu} ^{EM} =  g_{\mu\nu} L(F ) - L_F {F_{\nu}}^{\alpha} F_{\mu \alpha}\ ,
\yyy                     \label{SET-phi}
    T_{\mu\nu}^{\phi} =  2 \epsilon \d_\nu\phi\d_\mu\phi 
				    - g_{\mu\nu} \big (\epsilon \d^\alpha \phi \d_\alpha \phi - V(\phi)\big)\ .
\ear
Note that along this paper, we are considering just the contribution of an electric field, so that 
the only nonzero components of the electromagnetic field tensor are $F^{01}=-F^{10}$. Alternatively, one might consider the presence of only a magnetic field with the only nonzero components given by $F^{23}=-F^{32}$, 
as was done in a few previous papers 
  \cite{rodrigues23,rahul22,Bronnikov:2023aya,Lima:2023arg,canate22}.
Both cases can provide in general the same BB solutions with the appropriate though different Lagrangians (see \cite{Bronnikov:2022ofk}). Moreover, one may also consider the presence of 
both electric as magnetic fields (the dyonic case), but then the calculations turn out to be 
much more complex and do not lead to any new observations of interest. Finally, the presence 
of a canonical or phantom scalar field is required in order to satisfy the field equations, since 
otherwise the diagonal components of the Einstein tensor $G_\mu^\nu$ do not match 
the corresponding $T_\mu^\nu$ components of NED alone when considering BB solutions.

Note also that for static spherically/cylindrically symmetric space-times, as the ones analyzed below, \eq (\ref{eq-F}) for the electromagnetic field reduces to
\beq
\frac{1}{\sqrt{-g}}\frac{\d }{\d x^{1}}\left[\sqrt{-g}L_F F^{10}\right]=0\  ,
														\label{eq-F2}
\eeq
where $x^{1}$ is a radial coordinate. Then, it is straightforward to reconstruct the corresponding NED Lagrangian in terms of the radial coordinate for a particular metric of that kind. However, 
the reconstruction of the NED Lagrangian in terms of $F$ 
involving an electric field is subject to the possibility of inverting the expression 
of $F(x^1)$, which generally turns out to be difficult 
and not always unambiguous. In the following sections, we show explicitly the reconstruction procedure for a number of BB solutions
but in many cases express the Lagrangian $F$ in terms of the auxiliary invariant $P$ 
which is simply related to $x^1$.

\section{Spherically symmetric black bounces}                \label{Spherical}

Let us start by considering the class of static BB space-times with spherical symmetry. For simplicity, we focus on those metrics whose rr-component is the inverse of the tt-component. Then, the most general space-time metric matching such requirements is given by:
\beq              \label{ds-cy}
    		ds^2=f(r)dt^2 - f(r)^{-1}dr^2 - \Sigma^2(r)\left(d\theta^2 + \sin^2\theta d\varphi^2\right)\ .
\eeq
It is straightforward to obtain the corresponding components of the Einstein tensor that read:
\bearr                 \label{Gmn}
		G^0_0 = -\frac{f'(r) \Sigma '(r)}{\Sigma (r)}-\frac{f(r) \Sigma '(r)^2}{\Sigma (r)^2}-\frac{2 f(r) \Sigma ''(r)}{\Sigma
   (r)}+\frac{1}{\Sigma (r)^2}\ ,
\nnn 
		G^1_1 =-\frac{f'(r) \Sigma '(r)}{\Sigma (r)}-\frac{f(r) \Sigma '(r)^2}{\Sigma (r)^2}+\frac{1}{\Sigma (r)^2}\ ,				
\nnn
		G^2_2 = G^3_3 =-\frac{f'(r) \Sigma '(r)}{\Sigma (r)}-\frac{f''(r)}{2}-\frac{f(r) \Sigma ''(r)}{\Sigma (r)}\ .
\ear    
 Since the space-time is spherically symmetric, one has to assume that $\phi = \phi(r)$ while the electric field is also radial, with the only non-zero component
 given by $F^{01}= -F^{10}$. From the electrodynamics equations (\ref{eq-F2}), the electric field can be expressed as follows:
\beq
  \frac{1}{\sqrt{-g}}\frac{\d }{\d r}\left[\sqrt{-g}L_F F^{10}\right]=0\ 
  \quad \then \quad \Sigma^2(r)L_F F^{10}=q = \const,
\eeq
where $q$ represents a constant electric charge.  Then, the invariant $F$ can be obtained 
through an algebraic equation if the NED Lagrangian is known: 
\beq
    L_F^2  F = -\frac{q^2}{2 \Sigma ^4(r)}\ \quad \then \quad F=F(r)\ .
    \label{FrSpheSymm}
\eeq
  Furthermore, the stress-energy tensors (\ref{SET-F}) and (\ref{SET-phi}) are given by:
\bearr         \label{T-phi}
		{T^{\phi}}^{\mu}_{\ \nu} = \epsilon f\phi'{}^2 \diag (1, -1, 1, 1)
		 + \delta^\mu _{\nu} V(\phi)\ , 
\yyy           \label{T-F}		
		{T^{EM}}^{\mu}_{\ \nu} 
		=  \diag\Big(L+\frac{q^2}{L_F \Sigma ^4},\ L+\frac{q^2}{L_F \Sigma ^4},\ L ,\ L \Big)\ . 
\ear   
  As pointed out above, for these two tensors it holds 
  ${T^{EM}}^{0}_{\ 0} = {T^{EM}}^{1}_{\ 1}$ 
  and ${T^{\phi}}^{0}_{\ 0} = {T^{\phi}}^{2}_{\ 2}$, while the Einstein tensor in general
  does not have such symmetries, therefore, an arbitrary given metric cannot be ascribed 
  to a source in the form of a scalar field or an electric field only, but they do the job when 
  both come into play, as can be easily shown by considering a combination of both 
  stress-energy tensors: 
\bearr              \label{T02}
  		G^{2}_{\ 2} - G^{0}_{\ 0} = \kappa ^2T^{2}_{\ 2} - \kappa ^2T^{0}_{\ 0}
  		 = -\frac{\kappa ^2 q^2}{L_F \Sigma ^4}\ ,
\yyy                 \label{T01+}
    G^{1}_{\ 1} + G^{0}_{\ 0} = \kappa ^2T^{1}_{\ 1} + \kappa ^2T^{0}_{\ 0} 
          = 2 \kappa ^2 \left(L+\frac{q^2}{L_F \Sigma ^4}+V\right)\ .
\ear
   The scalar field equation \eqref{eq-phi} yields
\beq
	\frac{2}{\Sigma^2(r)}\partial_{r}\left[\Sigma^2(r)f(r)\partial_{r}\phi\right]=V'(\phi)\ .
			\label{eq-phi-2}
\eeq
  Furthermore, by combining again the ${0\choose 0}$ and ${1\choose 1}$ components of 
  the Einstein field equations \rf{eq-Ein} as
\beq              \label{T01}
		G^{1}_{\ 1} - G^{0}_{\ 0} = \frac{2 f \Sigma ''}{\Sigma} 
		= \kappa ^2T^{1}_{\ 1} - \kappa ^2T^{0}_{\ 0}=- 2 \kappa ^2 \epsilon  f \phi '^2\ ,
\eeq   
  the corresponding solution for the scalar field $\phi=\phi(r)$ is achieved. Hence, $L_F=L_F(r)$ 
  is provided as a function of $r$ through \eq\eqref{T02} whereas the corresponding scalar 
  potential $V=V(r)$ can be obtained through \eq\eqref{eq-phi-2}, and one can immediately 
  find the NED Lagrangian as $L=L(r)$ in (\ref{T01+}). However, as pointed above, for regular 
  solutions with an electric source, an explicit reconstruction of the expression for the NED 
  Lagrangian $L(F)$ is difficult analytically. To circumvent this issue, one might consider the 
  auxiliary field $P_{\mu\nu}=L_F F_{\mu\nu}$ and the Lagrangian as a function of 
\beq        \label{Pscalar}
  	        P = P^{\mu\nu}P_{\mu\nu}/4 = L_F^{2}F, 
\eeq  	
  where the invariant $P$ is given in terms of the electric charge by
\beq
         P=-\frac{q^2}{2 \Sigma^4}\ ,         \label{Pscalar1}
\eeq
which makes it easy to express $L$ in terms of $P$, if $L(r)$ is known.
  Meanwhile, if the function $F(P)$ is not monotonic (which is often the case), it is impossible 
  to obtain a unique expression for $L(F)$, and this Lagrangian function turns out to be 
  different in different space-time regions.
   
   In what follows, some particular examples of spherically symmetric BB solutions are analyzed,    
   and the corresponding Lagrangians are reconstructed.

\subsection{Simpson-Visser black bounce} 

The Simpson-Visser space-time is described by the metric \eqref{ds-cy}, where $f(r)$ and 
$\Sigma(r)$ are given by \cite{Simpson:2018tsi}
\beq
    f(r)=1-\frac{2m}{\sqrt{a^2+r^2}}, \quad \mbox{and} \quad \Sigma^2(r)=a^2+r^2\ .
    \label{SVspace-time}
\eeq
This space-time describes a BB as far as $a\neq 0$, since the radius of the 2-spheres is bounded from below due to $\Sigma>0$. The bounce might be hidden behind an event horizon if $a<2m$ while it describes a traversable wormhole in case of absence of event horizon, provided by $a>2m$. By following the procedure described above, the full set of functions describing the corresponding Lagrangians are obtained for the space-time 
\eqref{SVspace-time} as
\bearr
    \phi(r)=\frac{\tan ^{-1}\left(\frac{r}{a}\right)}{\kappa  \sqrt{-\epsilon }}\ ,
    		\qquad
    V(r)=\frac{4 m a^2}{5 \kappa ^2 \left(a^2+r^2\right)^{5/2}}\ ,
\nnn
    L(r)=-\frac{9 m a^2}{5 \kappa ^2 \left(a^2+r^2\right)^{5/2}}\ ,
    		\qquad
    L_F(r)=\frac{\kappa ^2 q^2\sqrt{a^2+r^2}}{3 a^2 m}\ .              \label{ReconsSV}
\ear
One can note that the scalar field and its potential $V$ coincide with those obtained in the case where the NED with a magnetic field was used \cite{rodrigues23}. However, the form of the NED Lagrangian is not the same as there, as one would expect since the two fields behave differently, and there
is no electric-magnetic duality in the general NED.

Using the above relations (\ref{ReconsSV}) together with \eqref{FrSpheSymm},  the scalar $F$ is obtained:
\beq
   		 F(r)=-\frac{9 m^2 a^4}{2q^2 \kappa ^4 \left(a^2+r^2\right)^3}\ .
\eeq
  It is now straightforward to assure the correctness of our calculations by verifying the
  validity of the  relation
\beq
                \frac{dL}{dr}  - L_F \frac{dF}{dr} = 0\ .                 \label{Cons}
\eeq

  In this particular case, the Lagrangian $L(F)$ can be explicitly expressed in terms of $F$ without the  
  need of using the auxiliary field $P$. The corresponding NED Lagrangian is
\begin{equation}
    L(F)=-\left(\frac{ 3\ 2^{5/2}\kappa ^{4} 3|q|^5}{125 a^4m^{2}}\right)^{1/3}|F|^{5/6}\ .
\end{equation}
  Moreover, the scalar field potential can be explicitly obtained in terms of $\phi$ by combining the expressions (\ref{ReconsSV}), leading to:
\beq
		V(\phi)=\frac{4 m}{5 \kappa ^2 a^3}\cos^5(\kappa\phi).         \label{VSPPhi}
\eeq
Thus the full Lagrangian is obtained for the case of Simpson-Visser BB. 

Let us also note that due to the expression $\Sigma(r) = \sqrt{a^2+r^2}$,
  we have $\Sigma''/\Sigma = a^2/\Sigma^4 > 0$, and in \rf{T01} we have to choose  
  $\epsilon=-1$, so that the scalar field is phantom, as should have been expected 
  because with this $\Sigma(r)$ the coordinate value $r=0$ corresponds either to a wormhole
  throat or to a BB in the precise meaning of these words.
  
In fact, we will get a similar situation in all examples to be considered in the present 
  paper: in other words, in all these BB space-times we are dealing with phantom 
  scalar fields. A more general and more complex situation is described in \cite{kb22-trap},
  involving scalar fields that are phantom only in a strong field region.

\subsection{Bardeen black bounce} 

Let us now consider an extension of the Bardeen solution BB, which is described by the following functions in the space-time metric \eqref{ds-cy},
\begin{equation}
    f(r)=1-\frac{2mr^2}{\left(a^2+r^2\right)^{3/2}}, \quad \mbox{and} \quad \Sigma^2(r)=a^2+r^2.
\end{equation}
Similarly to the previous case, for $a\neq0$ this class of space-time metrics describes either a BB hidden by two horizons when $0<a/m\leq4\sqrt{3}/9$ while it represents a traversable wormhole otherwise (for more details, see Ref.~\cite{lobo20}). Following again the same reconstruction procedure, by considering the full action (\ref{Action}), the corresponding magnitudes for the electromagnetic side and the scalar field potential yield:
\bear
    L(r)&=&\frac{3 m \left(34 a^4-91 a^2 r^2\right)}{35 \kappa ^2 \left(a^2+r^2\right)^{7/2}}\ ,\qquad
    L_F(r)=-\frac{q^2\kappa ^2 \left(a^2+r^2\right)^{3/2}}{a^2m \left(2 a^2-13 r^2\right)}\ ,
\nn
  \phi(r)&=&\frac 1\kappa \tan ^{-1}\left(\frac{r}{a}\right), \qq \epsilon =  -1, 
\qquad
   V(r)=\frac{4 m \left(7 a^2 r^2-8 a^4\right)}{35 \kappa ^2 \left(a^2+r^2\right)^{7/2}}\ . 
    \label{SolBardeen}
\ear
  The scalar field solution depends on the radial coordinate as in the previous example, but not 
  the scalar field potential, which
   is expressed in terms of the scalar field as
\beq
	V(\phi)=\frac{4 m \left(7 \tan^2(\kappa\phi)-8 \right)}{35 \kappa ^2 a^3}
					\cos^7(\kappa\phi).
			\label{VphiBardeen}
\eeq
  The electromagnetic scalar $F(r)$ is determined as
\begin{equation}
    F(r)=-\frac{a^4 \left(2 m a^2-13 m r^2\right)^2}{2q^2 \kappa ^4 \left(a^2+r^2\right)^5}\ .
\end{equation}
  Once again, as is necessary, the electromagnetic Lagrangian and the function $L_F$ 
  satisfy the condition \eqref{Cons}. 
  
  The function $F(r)$ cannot be inverted explicitly to obtain $r=r(F)$ and hence the corresponding 
  NED Lagrangian $L(F)$. However, we may use the auxiliary field $P$ to obtain $L(P)$, 
  which yields
\begin{equation}
    L(P)=-\frac{3\ 2^{3/4} m |P|^{5/4} \left(91 \sqrt{2} a^2 q-250 a^4 \sqrt{|P|}\right)}{35
   \kappa ^2 q^{7/2}}\ .
\end{equation}

These results show that BB solutions can be consistently reconstructed with an electric source, similarly to the cases with a magnetic source \cite{rahul22,canate22,rodrigues23}.

\begin{figure*}
\centering
\includegraphics[width=5.6cm]{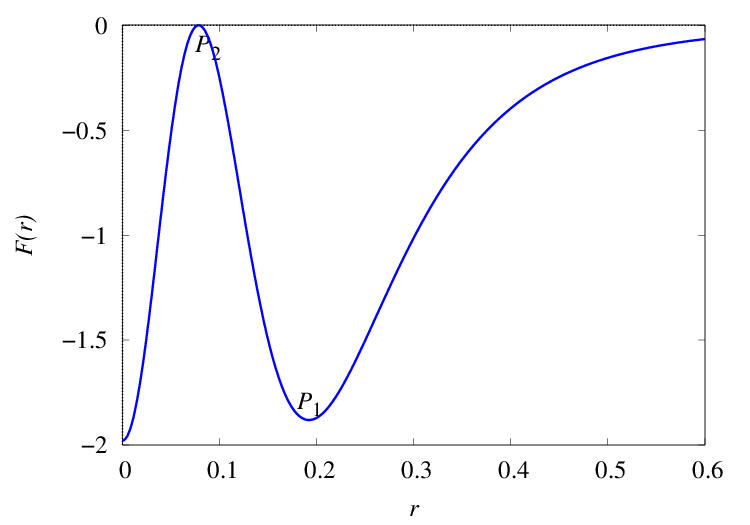} 
\includegraphics[width=5.6cm]{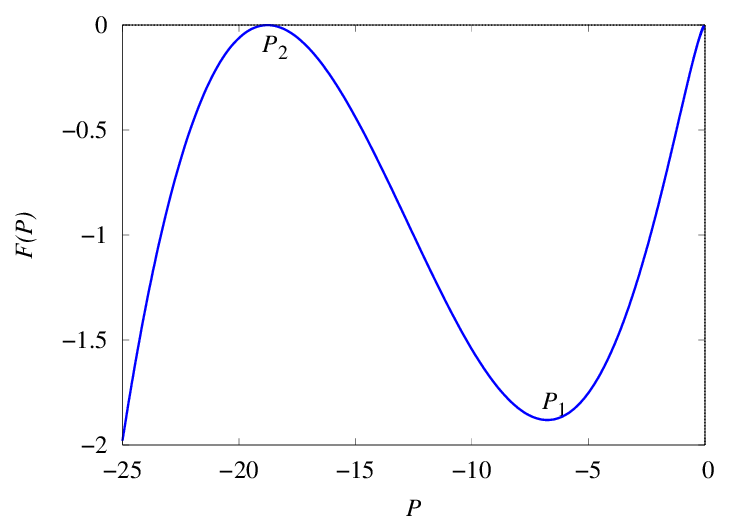} 
\includegraphics[width=5.6cm]{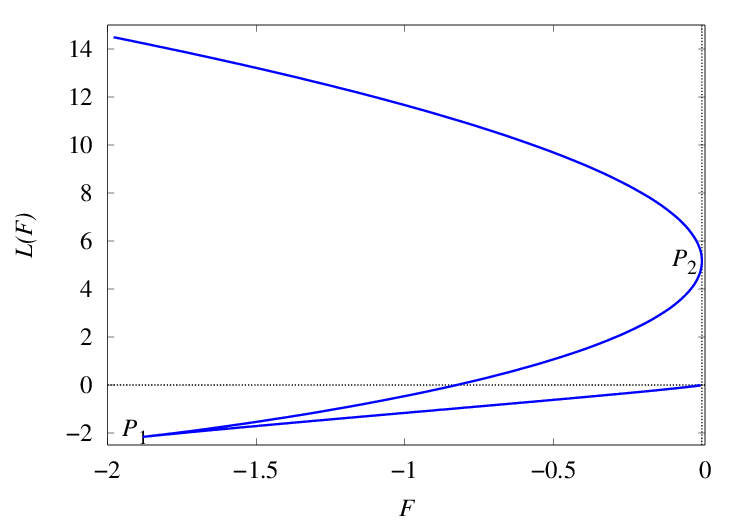}
\caption{\small 
		Plots of $F(r)$, $F(P)$ and $L(F)$ (left to right) for particular values of the parameters 
		of the Bardeen BB: $q = a =0.2$ and $m=1$. The extremum points $P_1$ and $P_2$
		of the function $F(P)$ correspond to branching of the Lagrangian function $L(F)$.}
		\label{fig-Bardeen}
\end{figure*}

  However, the absence of a monotonic relationship between the invariants $F$ and $P$ inevitably 
  leads to different branches of the Lagrangian function $L(F)$, each of these corresponding 
  to a monotonicity range of $F(P)$, as illustrated in Fig.\,\ref{fig-Bardeen}. Thus three different 
  NED theories are working in different ranges of $r$.

\section{Cylindrically symmetric black bounces}               \label{cyllindrical}

Let us now consider a class of BB solutions that are cylindrically symmetric instead of the spherically symmetric cases analysed in the previous section. Such space-times are usually known as Black Strings. For such solutions, a general space-time metric can be described by the following line element:
\begin{equation}
    ds^2=f(\rho)dt^2-f(\rho)^{-1}d\rho^2-\Sigma^2(\rho)(dz^2+d\varphi^2)\ .\label{Line_Cin}
\end{equation}
In this case, the nonzero components of the Einstein tensor are
\begin{eqnarray}
    G^{0}_{\ 0}&=&-\frac{f'(\rho ) \Sigma '(\rho )}{\Sigma (\rho )}-\frac{f(\rho ) \Sigma '(\rho )^2}{\Sigma (\rho )^2}-\frac{2
   f(\rho ) \Sigma ''(\rho )}{\Sigma (\rho )}\ ,\\
    G^{1}_{\ 1}&=&-\frac{f'(\rho ) \Sigma '(\rho )}{\Sigma (\rho )}-\frac{f(\rho ) \Sigma '(\rho )^2}{\Sigma (\rho )^2}\ ,\\
    G^{2}_{\ 2}&=&-\frac{f'(\rho ) \Sigma '(\rho )}{\Sigma (\rho )}-\frac{f''(\rho )}{2}-\frac{f(\rho ) \Sigma ''(\rho )}{\Sigma
   (\rho )}\ .
\end{eqnarray}
Meanwhile, the electric field has a similar expression to the one for the spherically symmetric case,
\begin{equation}
    F^{10}=\frac{q}{\Sigma^2(\rho)}L_F^{-1}\ .
\end{equation}
Note also that the relations \eqref{T02}-\eqref{T01} still hold in the cylindrical case with just changing $r\rightarrow \rho$, so that the procedure of reconstructing the full Lagrangian is 
similar to the one described in the previous section. In this section, two  particular cylindrically symmetric solutions are explored.

\subsection{Regular Black String I}

Let us consider the space-time \eqref{Line_Cin} with the following metric functions:
\begin{equation}
    f(\rho)=\frac{16k^4\left(1-\sqrt{a^2+\rho^2}\right)}{q^2\left(a^2+\rho^2\right)}, \qquad 
    		\Sigma^2(\rho)=\frac{q^2\left(\rho^2+a^2\right)}{4k^2}\ .
\end{equation}
  Here $k$ is a positive constant, $q$ is the effective electric/magnetic charge, and $a$ is again 
  a free regularization parameter leading to a bounce in the solution. This regular space-time (for 
  $a\neq0$) with cylindrical symmetry, also known as an ``inverted black hole'' \cite{kb79}
  has been previously analysed in Ref.~\cite{Bronnikov:2023aya}. Similarly to the cases above, it represents a regular solution that describes either a wormhole with the presence of two horizons ($a<1$) or a Bianchi type I solution with(out) an extremal horizon for $a=1$ ($a>1$) otherwise.  Using the relations  \eqref{T02}-\eqref{T01}, we obtain the following set of expressions for the regular black string:
\begin{eqnarray}
    \phi(\rho)&=&\frac 1\kappa \tan ^{-1}\left(\frac{\rho}{a}\right), \qq \epsilon =  -1,  
	\qquad
    V(\rho)=\frac{32 a^2 k^4 \left(3 \sqrt{a^2+\rho ^2}-5\right)}
    {15 \kappa ^2 q^2 \left(a^2+\rho ^2\right)^3}\ , 
\nn
    L(\rho)&=&-\frac{8 k^4 \left[a^2 \left(27 \sqrt{a^2+\rho ^2}-50\right)+30 \rho^2\right]}
    {15 \kappa ^2 q^2 \left(a^2+\rho^2\right)^3} ,
	\qquad
    L_F(\rho)=\frac{2 \kappa ^2 \left(a^2+\rho ^2\right)^{3/2}}{a^2 \left(3 \rho ^2-4 \sqrt{a^2+\rho ^2}\right)+4 \rho ^2
   \sqrt{a^2+\rho ^2}+3 a^4}\ .
\end{eqnarray}
  The scalar potential can be expressed explicitly in terms of $\phi$ as follows:
\beq
		V(\phi)=\frac{32 k^4 \left[3 a\sqrt{1+\tan^2(\kappa\phi)}-5\right]}
		{15 \kappa ^2 q^2 a^4} \cos^6(\kappa\phi) .
					\label{VPhiBSI}
\eeq

  Since the electromagnetic solution satisfies the condition \eqref{Cons}, the corresponding
  scalar $F$ can be obtained as
\begin{equation}
    F(\rho)=-\frac{2 k^4 \left[a^2 \left(3 \rho ^2-4 \sqrt{a^2+\rho ^2}\right)
    +4 \rho ^2 \sqrt{a^2+\rho ^2}+3  a^4\right]^2}
    {\kappa ^4 q^2 \left(a^2+\rho ^2\right)^5}\ .
\end{equation}
  This function cannot be inverted in order to obtain an analytic form for $L(F)$. Nevertheless, 
  as before, an explicit form of the NED Lagrangian in terms of the scalar $P=L_F^2F$, 
  which here has the same expression as in the spherically symmetric case \eqref{Pscalar}, yields
\begin{equation}
    L(P)=\frac{P}{30 \kappa ^2} \left[\frac{a^2}{k^2} 
    \left|2P\right|^{1/4}\left(27\left|q\right|^{1/2} k -40 |2P|^{1/4} q\right)+60\right]\ .
\end{equation}

\begin{figure*}[h]
\centering
\includegraphics[width=5.8cm]{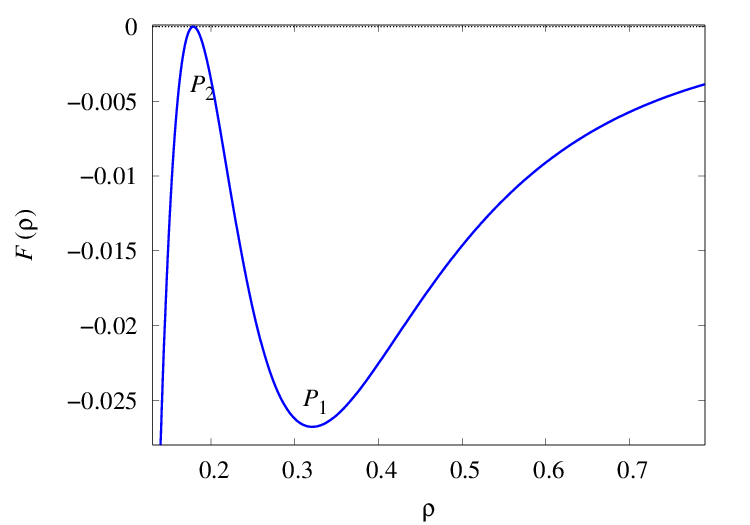} 
\includegraphics[width=5.8cm]{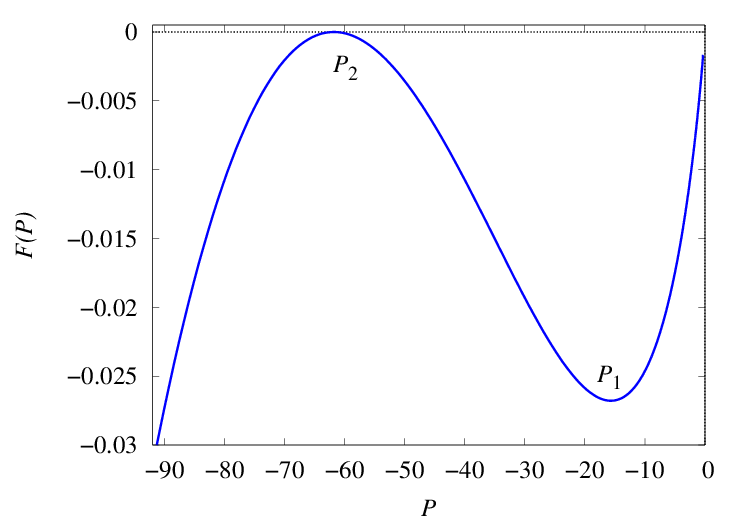}
\includegraphics[width=5.8cm]{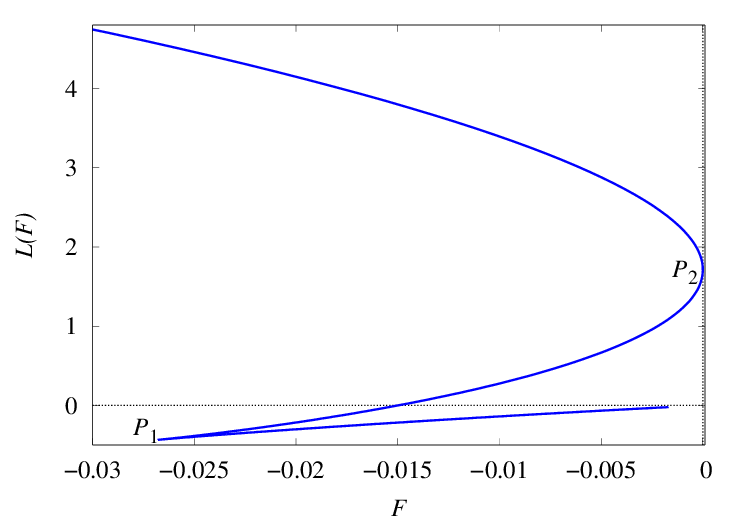}
\caption{\small 
		Plots of $F(\rho)$, $F(P)$ and $L(F)$ (left to right) for particular values of the parameters 
		of the black string-1 BB: $k = q = a = 0.2$. The qualitative behavior of all 
		functions is similar to Fig.\,1.}
		\label{fig-string1}
\end{figure*}

As was the case with Bardeen BB, the dependence $F(P)$ is not monotonic and
leads to three branches of the Lagrangian function $L(F)$ valid in different ranges of $\rho$, as illustrated in Fig.\,\ref{fig-string1}.

\subsection{Black string II}

As a second regular Black String solution, we consider the 
regularized version of Lemos's cylindrical solution \cite{Lemos95}, with the line element 
\eqref{Line_Cin} described by
\begin{equation}
    f(\rho)=\alpha^2 \left(a^2 +\rho ^2\right)-\frac{b}{\alpha  \sqrt{a^2+\rho ^2}}, 
    \qquad \Sigma^2(\rho)=a^2+\rho^2\ ,
    \label{BSII}
\end{equation}
  where $\alpha$, $b$ and $\alpha$ are positive constants. This space-time 
contains a regular BB at $\rho=0$ between two event horizons on its both sides 
  located at $\rho^2 = \rho^2_{h} =(b^{1/3}/\alpha)^2 - a^2$ as long as $a< b^{1/3}/\alpha$; 
  it has a horizon coinciding with a throat at $\rho =0$ (also called a ``black throat'') 
  if $a=b^{1/3}/\alpha$; lasty, it describes a traversable wormhole if $a >  b^{1/3}/\alpha$
  (for a complete analysis, see Ref.~\cite{Lima:2022pvc}).  Through the relations  
  \eqref{T02}--\eqref{T01}, the corresponding solution for the scalar field and the Lagrangian
  are reconstructed in terms of the radial coordinate $\rho$:
\begin{eqnarray}
    \phi(\rho)&=&  \frac 1 \kappa \tan^{-1}\left(\frac{\rho}{a}\right),\qq \epsilon = -1,
    \qquad
    V(\rho)=\frac{2 a^2 \left[5 \alpha ^3 (a^2+\rho ^2)^{3/2}+b\right]}
    {5 \alpha  \kappa ^2 (a^2+\rho^2)^{5/2}}\ ,
\nn
    L(\rho)&=&-\frac{3}{10 \alpha  \kappa ^2}\left(\frac{3 a^2 b}
    {\left(a^2+\rho^2\right)^{5/2}}+10 \alpha ^3\right),
\qquad
    L_F(\rho)=\frac{2 \alpha  \kappa ^2 q^2 \sqrt{a^2+\rho ^2}}{3 a^2 b}\ .
    \label{SolBSII}
\ear
The scalar field is the same as in the previous cases, while the electromagnetic magnitudes 
satisfy the condition \eqref{Cons}, and its Lagrangian is consistent with the one obtained in \cite{Lima:2023arg}. The corresponding scalar $F$ can be obtained through the condition \eqref{Cons}, leading to:
\begin{equation}
    F(\rho)=-\frac{9 a^4 b^2}{8 \alpha ^2 \kappa ^4 q^2 \left(a^2+\rho ^2\right)^3}\ .
\end{equation}
Contrary to the previous solution, here this function can be inverted in such a way that the NED Lagrangian can be written explicitly in terms of $F$ as follows:
\begin{equation}
    L(F)=-\frac{2 \sqrt{2} \sqrt[3]{3\alpha \kappa ^{4}}(-F)^{5/6} q^2 }{5 a^{4/3} 
    b^{2/3}\sqrt{q^{2/3}}}-\frac{3 \alpha ^2}{\kappa ^2}\ .
\end{equation}
Finally, the scalar field potential can be easily computed from \eqref{SolBSII},
\beq 
		V(\phi)=\frac{10 \alpha ^3 a^3 \sec^3(\kappa\phi) + 2b}
		{5 \alpha  \kappa ^2 a^3} \cos^5(\kappa\phi) .                        \label{VPhiBSIII}
\eeq
  The scalar field is phantom, as before. 
  
  In the next section, regular solutions in $2+1$ dimensions are reconstructed.

\section{Black bounces in 2+1 dimensions}  \label{21dimensions}

The line element that describes BB space-times in $2+1$ dimensions can be written as:
\begin{equation}
    ds^2=f(r)dt^2-f(r)^{-1}dr^2-\Sigma^2(r)d\varphi^2\ ,    \label{Line2Dgen}
\end{equation}
while the nonzero components of the Einsteins tensor are
\begin{eqnarray}
    G^{0}_{\ 0}&=&-\frac{f'(r) \Sigma '(r)}{2 \Sigma (r)}-\frac{f(r) \Sigma ''(r)}{\Sigma (r)}\ , \nonumber\\
    G^{1}_{\ 1}&=&-\frac{f'(r) \Sigma '(r)}{2 \Sigma (r)}\ ,\nonumber\\
    G^{2}_{\ 2}&=&-\frac{1}{2} f''(r)\ .
\end{eqnarray}
Once again one has in general $ G^{0}_{\ 0} \neq G^1_{\ 1}$ and $G^0_{\ 0} \neq G^{2}_{\ 2}$, which means that NED or a scalar field taken separately cannot be a source, but both together 
they do. As in the previous sections, here the only nonzero component of the electrodynamics 
tensor is assumed to be $F^{10}=-F^{01}$, so that there is only an electric field. Then, by solving the electrodynamics equations given in (\ref{eq-F2}), the electric field in $2+1$ dimensions 
takes the form
\begin{equation}
		    F^{10}=\frac{q^2}{\Sigma}L_F^{-1}\ ,
\end{equation}
whence the scalar $F$ is
\begin{equation}
    F=-\frac{q^2}{2 L_F^2 \Sigma^2}\ .
\end{equation}

By exploring the symmetries of the stress-energy tensor and by assuming the presence of a cosmological constant, one gets the following relations:\footnote
		{For this type of space-times in $2+1$ dimensions, a nonzero cosmological constant 
		is assumed since we are considering a regular version of the BTZ solution, which 
		corresponds to the Einstein equations in $2+1$ with a cosmological constant.}
\bearr              \label{T2D02}
  		G^{2}_{\ 2} - G^{0}_{\ 0} = \kappa ^2T^{2}_{\ 2} - \kappa ^2T^{0}_{\ 0} = -
  				\frac{\kappa ^2 q^2}{L_F \Sigma ^2}\ ,
\yyy                                  \label{T2D01+}
    G^{1}_{\ 1} + G^{0}_{\ 0} + \frac{2}{l^2} = \kappa ^2T^{1}_{\ 1} 
    + \kappa ^2T^{0}_{\ 0} = 2 \kappa ^2 \left(L+\frac{q^2}{L_F \Sigma ^2}+V\right)\ ,
\yyy 							    \label{T2D01-}
		G^{1}_{\ 1} - G^{0}_{\ 0} = \frac{f \Sigma ''}{\Sigma} 
		= \kappa ^2T^{1}_{\ 1} - \kappa ^2T^{0}_{\ 0}=-2 \kappa ^2 \epsilon  f \phi '^2\ . 
\ear
  These equations are very similar to those in $3+1$ dimensions. As far as $L_F$ is obtained as 
  a function of $r$, the corresponding NED Lagrangian $L(r)$ and the scalar field potential 
  $V(r)$ are computed using \eq \eqref{eq-phi}. We will analyze two explicit examples of
  such space-times.

\subsection{Regular BTZ Black Hole}

  The regular BTZ solution was proposed in Ref.~\cite{Furtado:2022tnb} 
  as a generalization of the Banados--Teitelboim--Zanelli black hole solution \cite{BTZ}
  and is described by the line element \eqref{Line2Dgen} with the functions
\begin{equation}
    f(r)=-M+\frac{r^2+a^2}{l^2},\quad \mbox{and} \quad \Sigma^2(r)=r^2+a^2\ ,
    \label{BTZmetric}
\end{equation}
   where $M$ is the dimensionless mass $a$ and $l$ are constants. This space-time is 
   singularity-free and contains event horizons if $a\leq l\sqrt{M}$ and represents a traversable 
   wormhole otherwise. By applying the reconstruction method described above, one finds:
\begin{eqnarray}
    L(r)&=& -\frac{3 a^2 M}{4 \kappa ^2 \left(a^2+r^2\right)^2}\ ,
    \qquad
    L_F(r)= \frac{\kappa ^2 q^2 \left(a^2+r^2\right)}{a^2 M}\ , 
\nn    
    V(r)&=& \frac{a^2 \left(2 a^2+l^2 M+2 r^2\right)}{4 \kappa ^2 l^2 (a^2+r^2)^2}\ , 
    \qquad
    \phi(r)= \frac{1}{\sqrt{2\kappa^2}} \tan ^{-1}\left(\frac{r}{a}\right), \qq \epsilon =-1.
    \label{BTZsolution}
\end{eqnarray}
  The electromagnetic invariant $F$ is found as
\beq
		 F(r)=-\frac{a^4 M^2}{2 \kappa ^4 q^2 \left(a^2+r^2\right)^3}\ .   \label{FBTZ}
\eeq

  Once again, we are dealing with a phantom scalar field, and using it together with 
  \rf{FBTZ}, $V(\phi)$ and $L(F)$ are computed explicitly as follows:
\begin{eqnarray}
    L(F)&=&-\frac{3 \kappa ^{2/3} \left(|F| q^2\right)^{2/3}}{2  (2a^2M)^{1/3} }\ ,
\nn
    V(\phi)&=&\frac{\cos ^2\left(\sqrt{2} \kappa  \phi \right) \left(4 a^2+l^2 M 
    \cos \left(2 \sqrt{2} \kappa  \phi \right)+l^2  M\right)}{8 a^2 \kappa ^2 l^2}\ .
   \label{VandLBTZ}
\end{eqnarray}
  One might choose to consider the auxiliary field $P^{\mu\nu}$ instead of the field 
  $F^{\mu\nu}$, then the NED Lagrangian is expressed as
\begin{equation}
		  L(P)=  -\frac{3 a^2 M P^2}{\kappa ^2 q^4}\ .
\end{equation}
  Thus the regular BTZ black hole is shown to be a consistent solution with NED in the presence 
  of a phantom scalar field.

\subsection{Regular Einstein-CIM space-time}

  Finally, another $2+1$-dimensional black hole solution is considered, described by a regular 
  version of the Einstein-CIM solution that arises from coupling of a three-dimensional gravitational 
  theory with conformally invariant Maxwell electrodynamics \cite{Hendi:2014mba}. 
  The original solution is given by the line element \eqref{Line2Dgen} with
\begin{equation}
    f(r)=-M+\frac{r^2}{l^2}-\frac{ Q }{2r}, \qquad \Sigma(r)=r\ ,\qq Q :=\left(2q^2\right)^{3/4},
\end{equation}
  where $M$ plays the role of an effective mass, and $q$ is the electric charge. After applying 
  the regularization method, these functions become:
\begin{equation}
    f(r)=-M+\frac{r^2+a^2}{l^2}-\frac{ Q }{2\sqrt{r^2+a^2}}, \qquad \Sigma^2(r)=r^2+a^2\ .
    \label{EinsCIMBlackHole}
\end{equation}
  Once again, we can apply the reconstruction formalism to determine the content that 
  generates this solution, which leads to
\begin{eqnarray}
    L(r) &=& \frac{10  Q  r^2-a^2 
    \left(15 M \sqrt{a^2+r^2}+14 Q \right)}{20 \kappa ^2 \left(a^2+r^2\right)^{5/2}}\ ,
\\
    L_F(r)&=& \frac{4 \kappa^2 q^2 \left(a^2+r^2\right)^{3/2}}
    {a^2 \left(4 M \sqrt{a^2+r^2}+3\,Q \right) - 3 Q  r^2}\ ,
\\
    F(r)&=&-\frac{\left[a^2 \left(4 M \sqrt{a^2+r^2}+3\,Q\right ) -3 Q r^2\right]^2}
    		{32 \kappa ^4 q^2 \left(a^2+r^2\right)^4}\ ,
\\
	\phi(r)&=& \frac{1}{\sqrt{2\kappa^2}} \tan ^{-1}\left(\frac{r}{a}\right), \qq \epsilon =-1.
\\
    V(r) &=& \frac{a^2 \left[l^2 \left(5 M \sqrt{a^2+r^2}+4 Q\right)+10 (a^2+r^2)^{3/2}\right]}	
    {20 \kappa ^2 l^2 \left(a^2+r^2\right)^{5/2}}.
\ear
  The form of the scalar field remains the same as in the previous cases, while the potential 
  of the scalar field shows a significant difference. For this solution, we are unable to write a 
  closed form for $L=L(F)$ since the expression $F=F(r)$ cannot be inverted, but the expression 
  for the scalar potential $V(\phi)$ can be computed,
\begin{equation}
    V(\phi)=\frac{1}{20 \kappa ^2}\Bigg(\frac{5 M \cos ^4\left(\sqrt{2} \kappa  \phi \right)}{a^2}
    +\frac{4  Q \cos ^6\left(\sqrt{2} \kappa  \phi\right) \sqrt{a^2 \sec^2\left(\sqrt{2} 
    \kappa  \phi\right)}}{a^4}+\frac{10 \cos ^2\left(\sqrt{2} \kappa  {\phi }\right)}{l^2}\Bigg)\ .
\end{equation}
  As before, using the auxiliary field $P = F L_F^2 = - \dfrac{q^2}{2(a^2+r^2)}$, one can 
  obtain the form of $L(P)$ as 
\begin{equation}
    L(P)=-\frac{3 a^2 M P^2}{\kappa ^2 q^4}-\frac{ P^{3/2} \left[48  a^2 P
			   (2q^2)^{1/4}+5  \left(2q^2\right)^{5/4}\right]}{5 \kappa ^2 q^4}\ .
\end{equation}
  Thus the regular $2+1$D space-time given by (\ref{EinsCIMBlackHole}) can also be reproduced 
  by this kind of source. 
It also turns out that in this case the dependence $F(P)$ is not monotonic, hence the 
  form of $L(F)$ is again multivalued, similarly to what was found in two previous examples, 
  see Fig.\,\ref{fig-ECIM}.
  
\begin{figure*}[h]
\centering
\includegraphics[width=5.5cm]{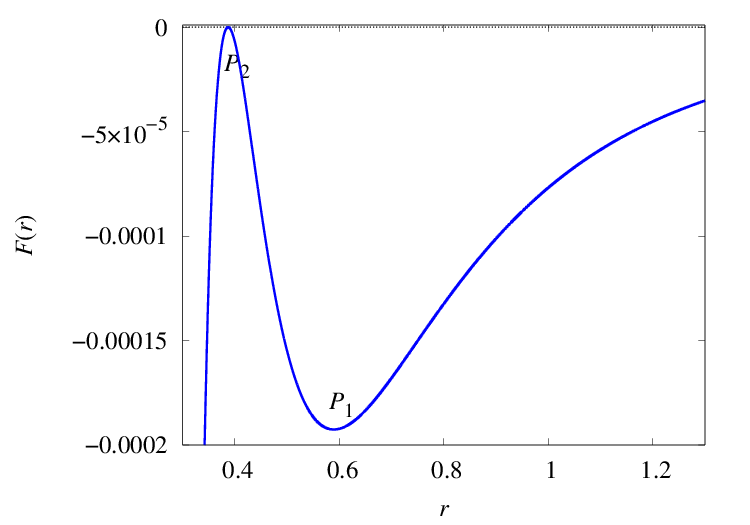} \quad
\includegraphics[width=5.5cm]{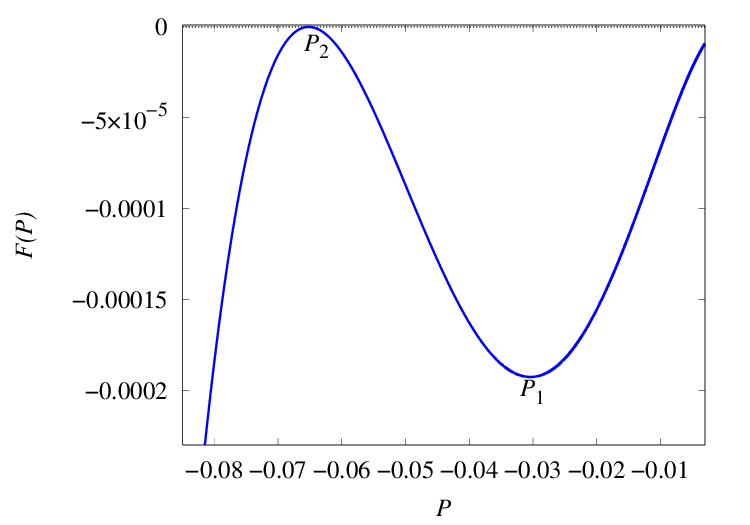} \quad
\includegraphics[width=5.5cm]{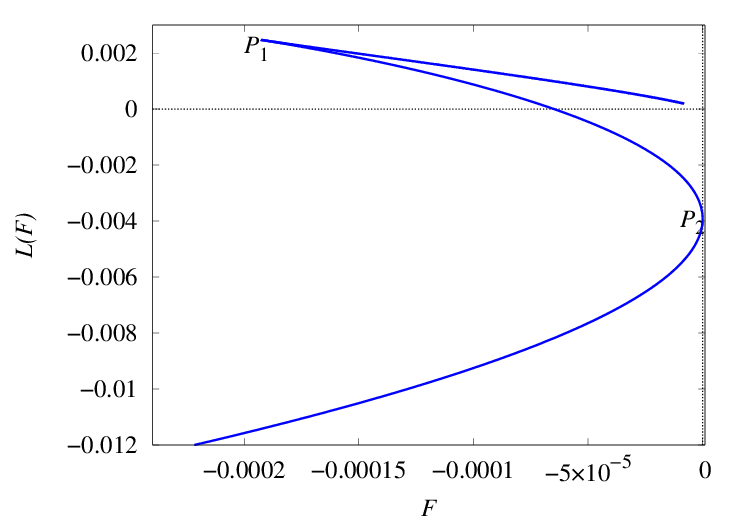}
\caption{\small 
		Plots of $F(r)$, $F(P)$ and $L(F)$ (left to right) for particular values of the parameters 
		of the regular Einstein-CIM BB: $a = Q =0.15$ and $M=1$. }
		\label{fig-ECIM}
\end{figure*}  
\section{Energy Conditions}                           \label{EN}

To conclude the paper, let us now analyze in a general way the energy conditions that can be 
applied to every case discussed above. To do so, we restrict ourselves to the Null Energy Condition
(NEC) since its violation implies violation of the the weak, strong, and dominant energy conditions
as well \cite{Visser:1995cc}. To analyze the NEC, one must compute 
\beq              \label{Enull}
	0 < E_{\rm null}\equiv \rho+p_{\|}= T^{1}_{\ 1} - T^{0}_{\ 0}
	=	\frac{1}{\kappa ^2}\left( G^{1}_{\ 1} - G^{0}_{\ 0}\right) 
	= \frac{2 f \Sigma ''}{\kappa ^2\Sigma} = - 2  \epsilon  f \phi '^2\ .
 \eeq
  Recall also that outside any outer horizon that may be present in a metric under consideration, 
  one has
\begin{equation}           \label{Tout}
    T^0{}_0=\rho,\qq   T^{1}{}_{1}=-p_{\|}, \qq T^{2}{}_{2}=T^{3}{}_{3}=p_{\bot}\ .
\end{equation}
  However, inside a simple horizon that may be present, the $t$ and $r$ coordinates swap 
  their timelike/spacelike nature, and the stress-energy tensor components do as well,
\begin{equation}\label{Toutt}
    T^0{}_0=-p_{\|},\qq      T^{1}{}_{1}=\rho,\qq     T^{2}{}_{2}=T^{3}{}_{3}=p_{\bot}\ .
\end{equation}
  With all that, the NEC can be written as
\begin{equation}              \label{EnullT}
	0 < E_{\rm null}=\begin{cases}
	      \ \ T^{0}_{\ 0}-T^{1}_{\ 1} ,\; f>0,  \\[5pt]
		    -T^{0}_{\ 0}+T^{1}_{\ 1} ,\;f<0,
	\end{cases}	
	\qq  \frac{1}{\kappa ^2}\left( G^{1}_{\ 1} - G^{0}_{\ 0}\right) = \frac{2 f \Sigma ''}{\kappa ^2\Sigma} = - 2  \epsilon  f \phi '^2\ ,
\end{equation}   
  while in all examples considered in this paper $\epsilon = -1$, and
\begin{equation}                                \label{EnullTT}
	E_{\rm null}=\left\{ \begin{array}{ll}
	     - 2   f \phi '^2 &,\; f>0\\[3pt]
		  \ \  2f \phi '^2&,\;f<0
	\end{array}     \right\}   = - 2 |f| \phi'{}^2.
\end{equation}   
 Since we here use an arbitrary $f$, \eq (\ref{EnullTT}) is valid for all the cases considered 
 previously. 
Equation (\ref{EnullTT}) also implies that a minimum of the radius $\Sigma(r)$ leads 
 to NEC violation when it takes place in any region. The same analysis was carried out, for example, 
 in \cite{Simpson:2018tsi} and \cite{kb18} using other notations and other sign conventions 
 but with the same result.
 
 Thus all regular black holes or wormholes accurately modeled by our space-time geometries
 (depending on the value of $a$) violate all classical energy conditions associated with the 
 stress-energy tensor.       

\section{Conclusions and discussion} \label{conclusions}

Along this paper, we have reviewed some of the most important BB space-times. Great attention drawn on this type of solutions lies on the feature that they are free from singularities, leading to different configurations depending on relative values of the free parameters of the metric tensor. 
An important achievement here is that a simple and straightforward method is implemented 
in order to reconstruct a particular theory that contains such BB space-times as exact solutions. 
To do so, in the framework of the Hilbert-Einstein action, a NED Lagrangian and a scalar field are involved, and it has been shown that every case leads to an analytical form of the Lagrangian for 
both the electromagnetic and scalar field, while the latter has always to be a phantom field. The inclusion of both NED and scalar field Lagrangians turns out to be completely necessary in order 
to satisfy the asymmetry among different components of the Einstein tensor. 
Such method is applied to a good number of solutions, including BB with spherical symmetry, 
black strings that possess cylindrical symmetry, and some BB solutions in $2+1$ dimensions. 

We have also considered, for the first time in the literature, electric sources of NED
for the solutions under consideration. 
A similar problem setting with magnetic sources has been
considered in \cite{rodrigues23,rahul22,Bronnikov:2023aya,Lima:2023arg,canate22} for
these and some others BB solutions. A possible regularization of spherical space-times more 
general than \rf{ds-cy}, with two arbitrary functions of the radial coordinate, was considered 
in \cite{kb22-trap}, it was shown that any such metric may be obtained as a solution of GR 
with a magnetic field in the framework of NED and a scalar field which can change its nature 
from canonical to phantom (a ``trapped ghost'' scalar \cite{kb10-trap}). It is of interest 
that an arbitrary static spherically symmetric metric can be obtained with a single field source
if one invokes the general (Bergmann-Wagoner-Nordtvedt) scalar-tensor theory of gravity, 
but this description turns out to be possible only piecewise \cite{kb23-stt}.

The main difference between electric and magnetic NED sources of BB metrics is that 
magnetic ones are always unambiguous while in electric ones the Lagrangian $L(F)$ suffers
branching at extremum points of the function $L(P)$. The same behavior of electric NED 
solutions was found to be inevitable in regular black hole models with a center $r=0$ at which
the electric field must vanish \cite{kb01-NED}. Unlike that, as we saw in the present paper, 
electric sources of BB solutions may avoid branching but possess this feature in many cases.
One can note that this branching, having much in common with phase transitions, seems to
be a phenomenon of separate interest, deserving a further study.


To know the form of the sources of BB and other regular space-times is fundamental since 
they influence the thermodynamics of black holes \cite{Rodrigues:2022qdp}, shadows 
\cite{dePaula:2023ozi}, and causality \cite{Tomizawa:2023vir}, and are essential for 
studying perturbations in the material content of the solutions \cite{Franzin:2023slm}. 

Hence, next steps might be focused on the analysis of the viability of such Lagrangians and 
the way such space-times can be extended to obtain more realistic theoretical models.

\section*{Acknowledgments}

G.A. and M.E.R. would like to thank Conselho Nacional de Desenvolvimento Cient\'ifico e 
Tecnol\'ogico - CNPq, Brazil  for partial financial support. G.A. and M.S. would like to thank 
Funda\c c\~ao Cearense de Apoio ao Desenvolvimento Cient\'ifico e Tecnol\'ogico (FUNCAP) 
for partial financial support. DSG is supported by the Spanish National Grant PID2020-117301GA-I00 funded by MCIN/AEI/10.13039/501100011033 (``ERDF A way of making Europe" and ``PGC Generaci\'on de Conocimiento"). K.B. acknowledges funding from the Ministry of Science
and Higher Education of the Russian Federation, Project ``New Phenomena in Particle Physics 
and the Early Universe'' FSWU-2023-0073.


\end{document}